# Dynamical Critical Behavior of Attractive Bose-Einstein Condensate Phase Transition

Yiping Chen[1], Munekazu Horikoshi[2,3], Kosuke Yoshioka[3] and Makoto Kuwata-Gonokami[1]

[1]*Department of Physics, Graduate School of Science, The University of Tokyo, 7-3-1 Hongo, Bunkyo-ku, Tokyo 113-0033, Japan*

[2]*Institute of Photon Science and Technology, Graduate School of Science, The University of Tokyo, 7-3-1 Hongo, Bunkyo-ku, Tokyo 113-8656, Japan*

[3]*Photon Science Center, Graduate School of Engineering, The University of Tokyo, 7-3-1 Hongo, Bunkyo-ku, Tokyo 113-8656, Japan*

When matter undergoes a continuous phase transition on a finite timescale, the Kibble-Zurek mechanism predicts universal scaling behavior with respect to structure formation. The scaling is dependent on the universality class and is irrelevant to the details of the system. Here, we examine this phenomenon by controlling the timescale of the phase transition to a Bose-Einstein condensate using sympathetic cooling of an ultracold Bose thermal cloud with tunable interactions in an elongated trap. The phase transition results in a diverse number of bright solitons and gray solitons in the condensate that undergo attractive and repulsive interactions, respectively. The power law dependence of the average soliton number on the timescale of the phase transition is measured for each interaction and compared. The results support the Kibble-Zurek mechanism, in that the scaling behavior is determined by universality and does not rely on the interaction properties.

Continuous phase transitions are ubiquitous phenomena that range from the generation of the early universe to the formation of multiferroics. Despite their diverse energy scales, continuous phase transitions are governed by some universal scaling laws that result from their statistical nature [1]. When continuous phase transitions occur on a finite timescale, they are also responsible for the defect and structure formation across the universe. The Kibble-Zurek mechanism (KZM) is a key theory that proposes universal power law scaling behavior for such processes [2-4]. Experiments that regard KZM defect formations have been conducted in recent years on various systems, including liquid crystals [5], superfluid $^3$He [6, 7], multiferroics [8] and ion chains [9, 10]. The KZM proves an experimentally simple and reliable criterion for the exploration of critical phenomena.

The process to Bose-Einstein condensate (BEC) from thermal states is also a type of continuous phase transition. Ultracold atomic gases are simple and highly controllable; therefore, they are ideal candidates for quantitative observations of critical phenomena. In most experiments, atomic BECs are produced with large repulsive interactions, typically around $100a_0$, to efficiently attain stable condensate states. Defects known as gray solitons (including planar solitons and solitonic vortices, this notation is applied for simplicity) or vortices can be spontaneously formed inside a repulsive BEC, depending on the specific geometry of the system. The dynamics and defect formation process of repulsive BEC phase transitions have been observed through several experiments that are in good agreement with the KZM [11-16].

On the other hand, BEC with attractive interaction in a harmonic trap is a self-focusing exotic object with limited atom number capacity [17]. An attractive BEC has the property to propagate without dispersion inside a waveguide and is thus called a bright matter-wave soliton [18]. The presence of multiple bright solitons in an elongated trap is typically referred to as a soliton train, which has been created by rapidly tuning the interaction strength of a preproduced repulsive

BEC to be weakly attractive via the Feshbach resonance [19-21]. Attractive BEC is considered a proper phase possessing the same order parameter as repulsive BEC that can be described by the Gross-Pitaevskii equation [22, 23]. This suggest that the phase transitions of attractive BEC and repulsive BEC are connected via universality, a widely believed empirical principle [24]. However, thus far, little has been studied regarding the spontaneous symmetry breaking process of an attractive BEC. Only a few experiments have successfully created an attractive BEC directly from a thermal cloud that contains hundreds of condensate atoms with relatively strong attractive interaction around $-30a_0$, and none of them is focused on this topic [25, 26]. A strong attraction decreases the number of condensate atoms and shortens the timescale of the dynamics, which is unfavorable for the observation of non-equilibrium critical phenomena. While weakening the interaction can increase the condensate atoms, evaporative cooling does not function efficiently in this regime for the thermal atoms to reach the condensate state.

We resolve this dilemma using the sympathetic cooling of weakly interacting bosons with a coolant. In this experiment, $^7$Li atoms and $^6$Li atoms are used as the boson specimen and coolant, respectively. The quench rate of bosons is varied by changing the total quench time. The spontaneous formation of multiple bright solitons is confirmed for faster quenches. A power law scaling is measured that describes the number of bright solitons formed after the BEC phase transition with different quench rates. By further repetition of the experiment in the weakly repulsive interacting region, we show that attractive BECs obey the same power law scaling as repulsive BECs, which is consistent with the KZM.

Near the critical temperature $T_c$ of a second-order phase transition, the equilibrium correlation length $\xi$ diverges as $\xi \sim |\varepsilon|^{-\nu}$, where $\varepsilon = (T_c - T)/T_c$ is the relative temperature and $\nu$ is the critical exponent for correlation. For the correlation to propagate to infinity, the corresponding relaxation time $\tau$ also diverges, $\tau \sim |\varepsilon|^{-z\nu}$, where $z$ is the critical exponent for relaxation. When $T_c$ is approached in a faster timescale than the relaxation time, the system cannot relax to equilibrium; therefore, disconnected regions exist during the phase transition.

An experimentally accessible parameter that well characterizes the critical phenomena is the number of defects that are spontaneously formed after the phase transition. The KZM proposes that the number of defects $N_d$ formed after the transition is determined by the average size of correlated domains $\hat{\xi}$ when the system becomes non-equilibrium. Repulsive BEC in a harmonic trap is an inhomogeneous system with a density-dependent local critical temperature. In such a case, the KZM predicts that defects created under a linear quench $\varepsilon(t) = t/\tau_Q$, where $\tau_Q$ is the quench rate, follows a simple power law scaling:

$$N_d \propto \tau_Q^{-(D-d)\frac{1+2\nu}{1+\nu z}}, \qquad (1)$$

where $D$ and $d$ are the dimensionality of the space and defects, respectively [4, 27].

On the other hand, the effect of finite-time phase transition on attractive BEC remains ambiguous. While the scaling laws of $\xi$ and $\tau$ should still hold and disconnected regions are bound to appear near $T_c$, the attraction preserves the shape of the condensate so that stable defects are not expected to be spontaneously formed as with other systems. Nonetheless, these regions do lead to observable consequences because the bright soliton is a localized structure that includes information about its phase domain. This means multiple-soliton states can be viewed as non-adiabatic excitations. If we assume that the number of bright solitons $N_b$ created after the phase transitions is also determined by the number of domains up to a factor, as illustrated in Fig. 1, we would also expect a KZM power law scaling as with repulsive BEC [28]. Accordingly, when linearly quenched through the critical point, we have:

$$N_b - 1 \propto \tau_Q^{-(D-d)\frac{1+2\nu}{1+\nu z}}, \qquad (2)$$

which is almost identical to Equation 1, except that a constant one is subtracted for that bright solitons correspond to condensates themselves. Here, the power law scaling of $N_b$ on $\tau_Q$ is measured to verify this scenario.

A detailed description of the apparatus employed can be found in Ref. [29]. The fermionic $^6$Li in two hyperfine states of $F=1/2$ serves as coolant for bosonic $^7$Li in the $|F=1, m_F=-1\rangle$ state. Our system is capable of sympathetically creating a mixture of superfluid fermions and bright solitons. However, the fermions are kept thermal in this experiment to avoid further complexities, such as immiscibility [30]. The mixed thermal cloud is first evaporatively cooled to ca. 300 nK. There are ca. $4\times10^5$ fermions in the trap with negligible atom loss throughout the subsequent processes. These fermions have a divergent scattering length, which allows the cloud to thermalize instantaneously. The typical number of atoms for the bosons at this moment is $4\times10^4$. The bosons have a negative scattering length of $a_{bb}=-3.2a_0$, which has been measured previously in Ref. [29]. The interaction between the latter is approximately $a_{bf}\sim 40a_0$ according to Ref. [31].

The atoms are then linearly quenched across the critical point of bosons. The final radial and axial trapping frequencies are $\omega_\rho=2\pi\times315$ Hz and $\omega_z=2\pi\times7$ Hz, respectively, with an aspect ratio of $\omega_\rho/\omega_z=45$. Under such conditions, the critical atom number $N_c$ that an individual bright soliton can support is $N_c=0.67\sqrt{\hbar/m\omega_\rho}/|a_{bb}|=8.5\times10^3$ [32]. The atoms are imaged using one of the two cameras. *In-situ* absorption images are acquired by a camera set horizontal to the axial direction of the trap. The imaging techniques employed are described in Ref. [33]. Alternatively, the trap and the magnetic field can be turned off to release the atoms and allow them free expend for about 4 ms before being imaged in the radial direction with the other camera. These time-of-flight (TOF) images are analyzed by bimodal fitting to estimate the temperatures.

For quantitative comparison, the exact power law $\tau_Q^{-\alpha}$ in Equation 1 and 2 must be evaluated for the present system. Critical exponents $v$ and $z$ are dependent only on the universality class of the phase transition. Both repulsive and attractive BEC are three-dimensional ($D=3$) and share the same order parameter; therefore, they belong to the same universality class, which is called the F-model. The critical exponents in this case are $\nu=0.67$ and $z=1.5$ [34]. The dimensionality $d$ can be either 1 or 2 in an elongated BEC, which involves some defect formation processes that are not clearly understood [16]. The quench rate $\tau_Q$ is controlled by fixing the initial temperature $T_i$ and the final temperature $T_f$ and changing the total quench time (evaporative cooling time) $t_q$. Thereby $t_q$ and $\tau_Q$ are proportional $\tau_Q = \left[T_c/(T_i-T_f)\right]t_q$ so that $\tau_Q^{-\alpha} \propto t_q^{-\alpha}$. These give a KZM predicted power law of $t_q^{-\alpha}$ with $\alpha_1 = 2.33$ for $d=1$ and $\alpha_2 = 1.17$ for $d=2$.

The evaporative cooling process is examined in Fig. 2. Since the bosons are weakly interacting, the maximum rate of their temperature drop is limited. The cooling efficiency is measured from TOF images as plotted in Fig. 2(a). Here we apply various $t_q$ by linearly decreasing the trap power for the same amount and we image the cloud after each quench. If thermally equilibrated, the cloud is expected to be cooled down from 300 to 120 nK at the end of these quenches. Only long evaporations satisfy such prerequisite and enable linear quenches with different $\tau_Q$. Figure 2(b) shows the temperature profiles for three quenches with long enough $t_q$'s, the temperatures are measured during these quenches and exhibit well linear property. From Fig. 2(c), the critical temperature is also estimated to be approximately $T_c = 190$ nK from the same data shown in Fig. 2(b). The condensate fraction grows smoothly during the quenches without violent collapse. These results show that this experiment is in the KZM configuration.

By using *in-situ* imaging, multiple bright solitons (soliton train) are detected near the center of the trap at the end of faster quenches, while a single bright soliton is produced for slower quenches, as shown in Fig. 2(d). The solitons are seeded from low density thermal atoms;

therefore, an individual soliton usually contains less than $N_c$ atoms and tends to contain more atoms with lesser number of solitons, in the range from 2000 to 8000 atoms. This indicates that major collapse cannot appear near the phase transition. The phase transition itself is stochastic, so that the resultant bright soliton number at the end of each quench typically varied from shot to shot, although faster quenches tended to produce more solitons. This behavior qualitatively matches the description of the KZM.

*In-situ* images are acquired at the end of each quench to measure the statistical feature of the bright solitons, with selected conditions where linear quenches end at the same final temperature. The initial and final temperature are measured to be 280 and 90 nK, respectively, for these data. Fifteen different $t_q$ are applied and 27 samples are taken for each. Condensates with more than twice the original thermal density peak are identified as bright solitons. The counts follow a Poisson distribution and the average bright soliton number versus the total quench time is shown in Fig. 3(a). The line in Fig. 3(a) corresponds to Equation 2 with a fitted power law with $\alpha = 2.57 \pm 0.46$ at a 95% confidence level. This value is in agreement with the predicted value of $\alpha_1 = 2.33$ from the KZM.

To confirm the validity of our finding, a similar series of experiments are conducted with weakly repulsive BEC. Here, the Feshbach magnetic field is used to tune the scattering length of the $|F=1, m_F=-1\rangle$ state $^7$Li to $a_{bb} \sim 4a_0$ and the cloud is precooled to 280 nK without alteration to the other properties of the system. Fifteen different total quench times $t_q$ are applied to cool the atoms through the critical temperature $T_c = 190 \text{ nK}$ down to 150 nK. The atoms are further cooled to 110 nK during a 4 s linear cooling and then hold in the trap for 1 s before imaging to increase the condensate fraction and enhance gray soliton growth. Here, *in-situ* imaging is used directly to count the gray solitons without expanding the gas, because the weakly interacting feature secures a higher atom density so that defects are easily recognizable.

Density depletions of more than 20% are counted as gray solitons. Twenty four samples are taken for each quench rate, and the counting results are given in Fig. 3(b). The line in Fig. 3(b) represents a power law fitting according to Equation 1 with $\alpha = 2.12 \pm 0.44$ at a 95% confidence level. This value is again very close to the value predicted by the KZM, $\alpha_1 = 2.33$.

The result for gray solitons is consistent with the result of Ref. [16]. The healing length $\xi_l$ of the present gray solitons is estimated to be approximately $\xi_l = 1/\sqrt{8\pi a_{bb} n_{con}} \sim 10$ μm, where $n_{con}$ is the condensate density, and the domain size predicted by the KZM is on the order of about 1 μm. Both are smaller than the axial harmonic oscillator length of 13 μm. Hence, defect formation in the present system corresponds to the high aspect ratio condition in Ref. [16]. As determined from the measured power law with $\alpha = 2.12 \pm 0.44$, we conclude that the dimensionality of defect formation in the system is 1. This also means the gray soliton created in our system is in fact a solitonic vortex. In the attractive BEC case, the growth condition of the condensate is similar to the repulsive case. From the measured power law with $\alpha = 2.57 \pm 0.46$, it is confirmed that the KZM explains the growth process of bright solitons as analogous with gray soliton formation in repulsive BEC.

A $\tau_Q$ range for the solitons to halve in number is examined in this experiment; therefore, the accuracy is adequate to show the consistency with KZM and to compare the critical phenomena between the attractive and repulsive case. In this experiment, both the bright solitons and gray solitons are considerably stable and can survive in the trap for seconds after the quench, which is sufficiently long to neglect the possible difference in growth time introduced by the variation of $t_q$, similar to that already discussed in Ref. [12]. The ground state single bright soliton is stable, which can be held in the trap for tens of seconds.

The weakly interacting property results in some unique features. The timescale of this experiment is about 1 order of magnitude longer compared to other finite time phase transition experiments using ultracold atoms. This can be explained by a long scattering time. Recall that $a_{bb} \sim 4a_0$, the scattering time $\tau_0 = \left(8\pi n_b a_{bb}^2 v_{mean}\right)^{-1} \sim 50 \text{ s}$, where $n_b$ is the thermal density of bosons and $v_{mean}$ is the Maxwell mean velocity, is comparable with the timescale of the BEC phase transition. The major downside of the weakly interacting property is the low cooling efficiency, which limits the range of quench time and restricted the access to a faster cooling regime. The dominant force for the bosons to cool down is their interaction with the fermions, which is about $a_{bf} \sim 40a_0$ [31]. The scattering time between the two species is hence $\tau_{bf} = \left(8\pi n_{bf} a_{bf}^2 v_{mean}\right)^{-1} \sim 200 \text{ ms}$, where $n_{bf}$ is the average thermal density of the mixture. This long scattering time is responsible for the low cooling rate observed in this experiment.

In conclusion, we have observed the spontaneous creation of bright solitons and gray solitons in the BEC phase transition using ultracold $^7$Li atomic gas with weakly attractive and repulsive interaction, respectively, via sympathetic cooling. The results show that soliton numbers as a function of the quench time follows power-law scaling that is consist with the Kibble-Zurek mechanism in both the weakly attractive and repulsive regime. This paper is the first experiment to compare the critical phenomena of repulsive and attractive BEC. The result implies that the phase transition and critical dynamics remain robust against quantum instabilities and are irrelevant to microscopic details of the interaction. This work also provides experimental information for the investigation of finite temperature attractive BECs and their evolutions. Future works may include the study of finite temperature attractive BEC and the critical properties of an ideal BEC.

The present study was supported by a Grant-in-Aid for Scientific Research on Innovative Areas (No. 18H05406).

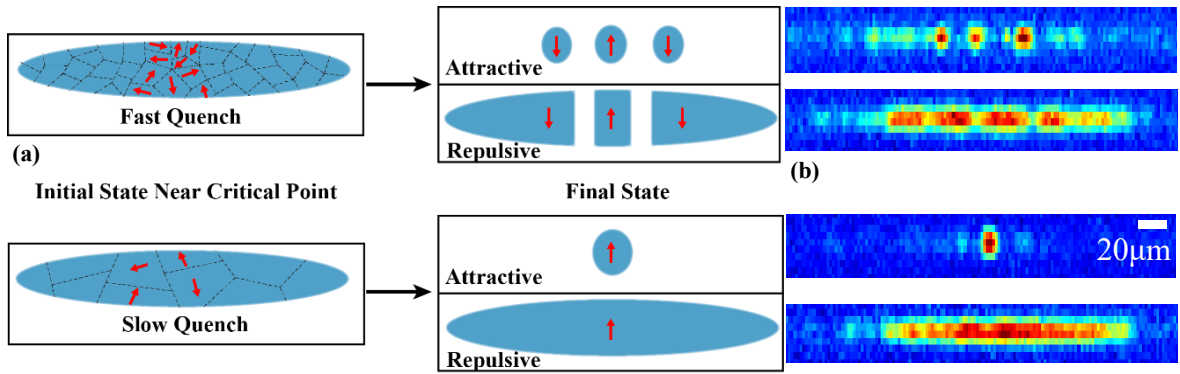

FIG. 1. Critical behavior of BEC in an elongated trap. (a) The arrows indicate the phases of the wave function. The average size of independent phase domains present near the critical point is determined by the quench rate. In the repulsive case, gray solitons appear between the domains, while in the attractive case, independent BEC domains themselves become bright solitons. (b) *In-situ* samples of finite temperature attractive BEC and repulsive BEC of $^7$Li after phase transitions. Multiple bright solitons or gray solitons are produced after fast quenches.

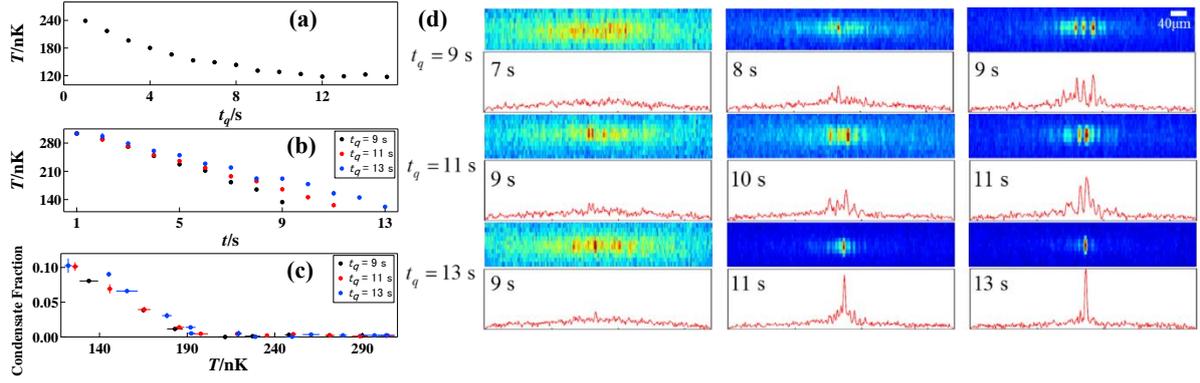

FIG. 2. Properties of attractively interacting bosons during quenches. (a) Cooling efficiency measured from TOF imaging. The cloud reaches a final temperature of 120 nK only for evaporations longer than 9 s, indicating a maximum cooling rate of about 20 nK/s. (b) Temperature profiles examined from TOF imaging. The quenches start at 300 nK and stop at 120 nK with different total quench times. Linear decreases in temperature can be observed. (c) Condensate fractions. A critical temperature of 190 nK is measured. The condensate fractions increase similarly for all quenches after the phase transition. (d) Sample *in-situ* images and their optical densities. The first row corresponds to atoms merely below the critical temperature. The second row presents growing solitons. The third row shows three, two and one solitons (from top to bottom) formed at the end of linear quenches.

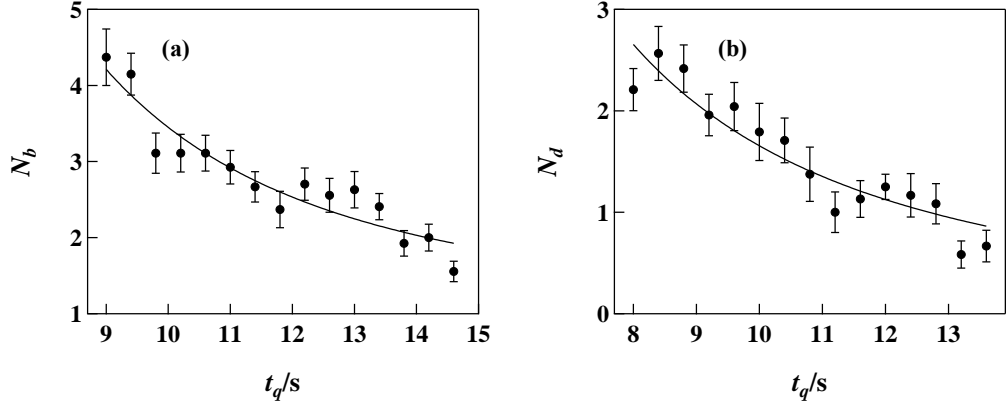

FIG. 3. Counting statistics of spontaneously formed solitons under different quench times. (a) Bright solitons with attractive interaction. The line shows the fitted power law with the exponent $2.12 \pm 0.44$. Averaged over 27 data. (b) Gray solitons in repulsive BEC. The line shows the fitted power law with the exponent $2.57 \pm 0.46$. Averaged over 24 data. The error bars show the standard error of the mean.